\documentclass[journal]{IEEEtran}
\usepackage{amsmath,amsfonts}
\usepackage{amssymb}
\usepackage{amsthm}
\usepackage{cite}
\usepackage{color}
\usepackage{xcolor}
\usepackage{epsfig,latexsym}
\usepackage{float}
\usepackage{fancyhdr}
\usepackage{graphicx}
\usepackage{indentfirst}
\usepackage{mdwmath}
\usepackage{mdwtab}
\usepackage{setspace}
\usepackage{textcomp}
\usepackage{times}
\usepackage{url}
\usepackage{multirow}
\usepackage{verbatim}
\usepackage{algorithmic}
\usepackage{balance}

\newtheorem{remark}{Remark}
\newtheorem{theorem}{Theorem}

\newtheorem{lemma}{Lemma}

\newtheorem{corollary}{Corollary}

\newtheorem{proposition}{Proposition}

\begin{document}

\title{On the Performance of Pinching-Antenna Systems (PASS) Under Dynamic Channels with Blockages}
\author{Jinhua Wang, Jun Wang, Tianwei Hou,~\IEEEmembership{Member,~IEEE}, Anna Li,~\IEEEmembership{Member,~IEEE}, \\
Yuanwei Liu,~\IEEEmembership{Fellow,~IEEE}, and Arumugam Nallanathan,~\IEEEmembership{Fellow,~IEEE}

\thanks{This work is supported in part by the Beijing Natural Science Foundation L232041. (Corresponding author: Tianwei Hou.)}
\thanks{Jinhua Wang, Jun Wang, and Tianwei Hou are with the School of Electronic and Information Engineering, Beijing Jiaotong University, Beijing 100044, China (e-mail: 25110041@bjtu.edu.cn, wangjun1@bjtu.edu.cn, and twhou@bjtu.edu.cn).}
\thanks{Anna Li is with the School of Computing and Communications, Lancaster University, Lancaster LA1 4WA, U.K. (e-mail: a.li16@lancaster.ac.uk).}
\thanks{Yuanwei Liu is with the Department of Electrical and Electronic Engineering, The University of Hong Kong, Hong Kong (e-mail: yuanwei@hku.hk).}
\thanks{Arumugam Nallanathan is with the School of Electronic Engineering and Computer Science, Queen Mary University of London, London E1 4NS, U.K., and also with the Department of Electronic Engineering, Kyung Hee University, Yongin-si, Gyeonggi-do 17104, Korea. (e-mail: a.nallanathan@qmul.ac.uk).}
}

\maketitle

\begin{abstract}
The performance of pinching-antenna systems (PASS) is fundamentally affected by line-of-sight (LoS) blockage in practical environments. In this paper, PASS is investigated under realistic, obstacle-induced blockage by jointly considering the LoS and non-LoS (NLoS) components, rather than relying on a LoS channel or a probabilistic blockage model. A geometry-aware blockage model is adopted, where a blockage region on the waveguide is defined according to the actual locations and geometric features of obstacles, such that a pinching-antenna (PA) located within the blockage region is unable to establish a LoS link to the user equipment (UE). The channel models of PASS are developed by jointly accounting for in-waveguide attenuation and spatial propagation loss. To quantify the impact of channel factors on PASS performance, a single-PA single-UE scenario is studied under Rayleigh and Rician fading channels. Closed-form expressions for the outage probability are derived for both cases. For the ergodic rate, a closed-form expression is obtained in the Rayleigh case, while a complete analytical expression and an approximate closed-form expression are derived in the Rician case. Analytical expressions are derived for the endpoints of the blockage region, and the deployment criteria of optimal PA are provided. Simulation results validate the analysis and reveal that: i) NLoS scattering has a twofold effect on PASS performance, potentially degrading the outage performance while improving the rate performance under Rician fading; ii) Sufficiently strong NLoS scattering can still sustain communication in the presence of LoS blockage; iii) The optimal PA position is jointly determined by the environment geometry and the interplay between spatial propagation loss and in-waveguide attenuation.
\end{abstract}


\section{Introduction}

Multiple antenna techniques have long been recognized as a fundamental means to improve the reliability, spectral efficiency, and spatial degrees of freedom of wireless communications~\cite{6736761}. Conventional antenna arrays can effectively improve performance in many scenarios by exploiting spatial diversity and multiplexing gains. However, once deployed, the position of antenna is fixed, which limits their ability to adapt to environmental dynamics, such as line-of-sight (LoS) blockage and user equipment (UE) mobility. As a result, conventional antenna arrays may suffer considerable performance degradation in challenging propagation environments, since they cannot reconstruct a favorable channel condition.
A variety of reconfigurable and flexible antenna technologies have been proposed, which can exploit small-scale fading and convert it to performance gains in dynamic environments, particularly under non-LoS (NLoS) conditions~\cite{10858129}. Reconfigurable intelligent surfaces (RISs) can create equivalent LoS links by dynamically reflecting or transmitting signals to bypass obstacles~\cite{9424177}, whereas fluid antennas and movable antennas enhance spatial diversity by physically or electronically shifting antenna positions within a region of a few wavelengths~\cite{10286328, 9264694}. Although such techniques are effective at mitigating local fading fluctuations, they are not inherent to combat large-scale fading, which is dominated by propagation geometry, path loss, and LoS availability.

The recently proposed pinching-antenna systems (PASS) offer a new form of flexible-antenna architecture~\cite{11169486, 11434944}. PASS consists of dielectric waveguides and multiple separate dielectric pinching-antennas (PAs) as radiation points~\cite{suzuki2022pinching, example2025}. By flexibly activating PAs along a wide spatial range of waveguide, PASS can establish favorable LoS links, which enables adaptive beamforming and spatial diversity gains for UE located in complex environments. Due to its distinctive advantages, PASS has received growing research attention.
The system models and prospects of PASS are initially established in~\cite{ding2025flexible}, which also provides broad insights into multiple key research directions, including the applications of non-orthogonal multiple access and multiple-input single-output. The fundamental performance analysis of a basic PASS, which consists of a single waveguide and a single PA, is investigated in~\cite{10976621}. The uplink and downlink performance of PASS with multiple PAs are studied in~\cite{11195810} and~\cite{10896748}, respectively, which also provide strategies for multi-PA activation. The principles of joint transmit and beamforming optimization of PAs are investigated in~\cite{11202577}, where a physics-based coupler model of PASS is proposed. By investigating a downlink PASS with PAs clustered on an array in~\cite{11434513}, the realization of PASS is explored. The capacity limits of multi-user uplink and downlink channels for multi-PA cases, as well as the inner and outer bounds, are analyzed in~\cite{11358830}, which provides a theoretical baseline for evaluating multiple access schemes. The in-waveguide attenuation is explicitly incorporated into PASS modeling and optimization in~\cite{xu2025pinchingantennasystemsinwaveguideattenuation}, where the determination standard of optimal PA position is concluded by categorically discussing the tradeoff between free-space path loss and in-waveguide attenuation.

A key premise behind the advantages of PASS is the availability of a favorable LoS path between the activated PA and the UE, which greatly expands the range of applications for PASS, such as indoor positioning~\cite{zhang2025pinchingantennasystemspassbasedindoor} and integrated sensing and communications~\cite{11424157}. However, the LoS path may be blocked by surrounding obstacles, such as walls, furniture, and human bodies. Once blockage occurs, the expected benefit of selecting an optimal PA position to establish a LoS link may be significantly reduced or eliminated. Therefore, it is vital to properly address the challenge of LoS blockage before the practical benefits of PASS can be fully realized.

Several studies examine the impact of LoS blockage on PASS and provide unique insights, most of which have adopt simplified probabilistic LoS blockage models for analytical tractability. For instance, the LoS probability is typically expressed as a simple exponential function parameterized by a LoS blockage exponent~\cite{7448962, 3gpp.38.901}. While LoS blockage degrades performance in single-user scenarios, it may actually benefit multi-user systems by suppressing co-channel interference~\cite{11036558}. To exploit LoS blockage for interference suppression, the problem of joint waveguide and PA association is studied in~\cite{11178241}, where the size and position of obstacles are perfectly known. Under a low-loss waveguide setting considered in~\cite{xu2025pinchingantennadesignlosblockage}, the in-waveguide attenuation is evaluated to have a negligible impact compared to blockage under meter-level propagation, particularly in densely obstructed environments. By exploiting LoS blockage as a multiple-access resource rather than treating it as a purely detrimental effect, a type of technique termed environment division multiple access is proposed for PASS in~\cite{ding2025environmentdivisionmultipleaccess}. Both LoS blockage and inter-cell interference are considered in~\cite{11315149}, where spatially distributed waveguides are connected to different base stations. Building on the cylindrical blockage model, both discrete and continuous PA placement designs are studied in~\cite{xie2026pinchingantennasblockageawareenvironments}, thereby treating blockage-aware LoS transitions as an integral part of PASS modeling and optimization. A cube-shaped blockage model for PASS is considered in~\cite{zhao2026robustsecureblockageawarepinching}, which maintains LoS connectivity for legitimate users while leveraging blockage to disrupt eavesdroppers’ channels under imperfect channel state information.

In general, with the meter-scale activation range of PAs, PASS can create LoS propagation conditions that are a hundred times stronger than NLoS scattering, which means that the contribution of NLoS components to the overall performance of PASS is negligible~\cite{yang2025pinching}. However, it is worth emphasizing that NLoS paths could be significant in extreme and obstacle-rich scenarios, such as indoor environments, power cable trenches, and underground utility tunnels, which suggests that NLoS scattering deserves consideration in PASS modeling, configuration, or beam training~\cite{11364174}.
Several studies have investigated PASS in the presence of NLoS propagation and made fundamental contributions. A tractable channel model for PASS that includes NLoS components is developed in~\cite{xu2025pinchingantennadesignrandomlos}, which combines a distance-dependent probabilistic LoS component with cluster-based random NLoS scattering. By formulating both average signal-to-noise (SNR) threshold coverage maximization and fairness-aware worst-grid average-SNR maximization, a blockage-aware network optimization framework for PASS is investigated within hundred-meter-scale areas in~\cite{xu2026environmentawarenetworkleveldesigngeneralized}, which reveals that maximizing spatial average SNR alone may cause low performance in blockage-affected regions.

\subsection{Motivation and Contribution}

From the above discussion, a following fundamental question naturally arises: how does PASS perform in obstacle-rich environments when both LoS blockage effect and NLoS propagation are jointly taken into account? To solve this problem, we aim to reveal the joint affection of the PA, obstacles, and UE on the performance of PASS in obstacle-rich environments. Specifically, we establish a geometry-aware framework that directly links obstacle-induced blockage to the PASS channel, rather than using LoS probability. By explicitly incorporating both LoS and NLoS components into the spatial channel model, the performance of PASS with realistic, obstacle-induced blockage can be analyzed under complete propagation conditions, which differs from existing LoS channel or probabilistic LoS-based models. The main contributions of this paper are summarized as follows:
\begin{itemize}
  \item To explicitly links the PASS channel to realistic obstacle-induced blockage, a geometry-aware blockage model is adopted. According to the locations and geometric features of obstacles, a blockage region is defined on the waveguide. The channel and signal models that jointly account for in-waveguide and spatial channel gains are developed.
  \item To quantify the impact of channel factors on the performance of PASS, a single-PA single-UE scenario is considered under both Rayleigh and Rician fading channels. Subsequently, closed-form expressions are derived for the outage probability and ergodic rate under Rayleigh fading, as well as for the outage probability under Rician fading. For the Rician ergodic rate, a tractable analytical expression is derived together with an accurate closed-form approximation. Moreover, the ergodic rate under the Rician channel is shown to strictly exceed that under the LoS channel.
  \item Analytical expressions for the endpoints of the blockage region are derived. Furthermore, in the Rayleigh case, a closed-form solution for the optimal PA position is derived. In the Rician case, maximizing received SNR is provided as an effective criterion for determining the optimal PA position, thereby avoiding the use of complicated optimization algorithms.
  \item Simulation results validate our analysis and further reveal several useful insights: 1) NLoS scattering has a twofold effect, degrading the outage performance while improving the ergodic rate of PASS under Rician fading; 2) In the absence of the LoS path, i.e., in the Rayleigh case, sufficiently strong NLoS scattering can still support communication requirement; 3) The optimal PA position depends on the relative locations of the waveguide, obstacles, and UE, as well as the interplay between in-waveguide attenuation and spatial propagation loss.
\end{itemize}

\subsection{Organization and Notations}
The rest of this paper is organized as follows. Section \uppercase\expandafter{\romannumeral2} introduces the antenna model, blockage model, and the channel and signal model. Section \uppercase\expandafter{\romannumeral3} analyzes the performance of PASS under both Rayleigh and Rician fading channels, and also presents solutions for the optimal PA position. Section \uppercase\expandafter{\romannumeral4} provides the numerical results and performance discussion, and Section \uppercase\expandafter{\romannumeral5} concludes the paper. $\mathbb{E}\left[ \cdot \right]$ denotes the mathematical expectation, and $\mathbb{P}\left( \cdot \right)$ denotes the probability. $X \sim \mathcal{CN}\left( \mu, \sigma^2 \right)$ denotes a complex Gaussian contribution with a mean of $\mu$ and a variance of $\sigma^2$.

\section{System Model}

\begin{figure}[t!]
  \centering
  \includegraphics[width = 3.4 in]{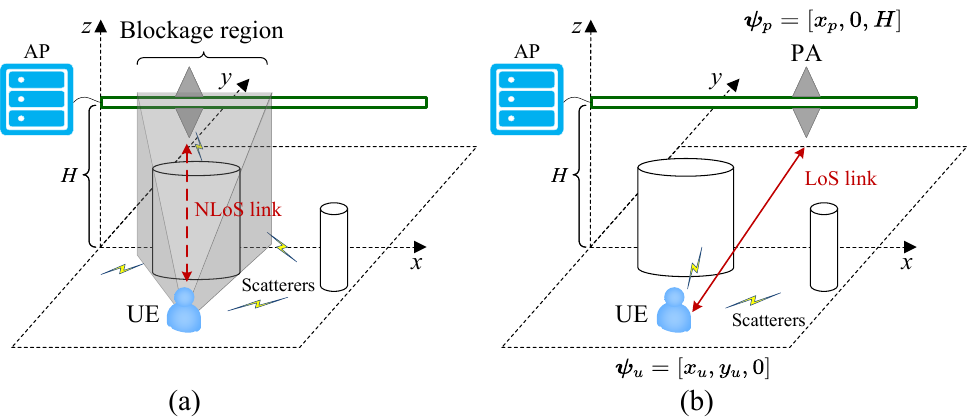}
  \caption{PASS in a blockage scenario under dynamic channels: (a) under a Rayleigh fading channel, and (b) under a Rician fading channel.}\label{fig: PASS under dynamic channels}
\end{figure}

\subsection{Pinching-Antenna Model}
A rectangular service area is defined as
\begin{equation}\label{service area def}
\mathcal{A} \triangleq \left\{ \left( {x,y} \right) \in {\mathbb{R}^2}\,\middle|\,0 \leqslant x \leqslant {D_L}, - \frac{{{D_W}}}{2} \leqslant y \leqslant \frac{D_W}{2} \right\},
\end{equation}
where $D_L$ and $D_W$ denote the length and width of the service area, respectively. The center of the service area serves as the origin of the three-dimensional Cartesian coordinate system.

A downlink PASS equipped with a single waveguide is considered. The waveguide is placed parallel to the $x$-axis, with its location defined as $\left( \left[ 0, D_L \right], 0, H \right)$, where $H$ denotes the height of the waveguide. The access point (AP) is located over the origin, i.e., $\boldsymbol{\psi} _0 = \left( 0, 0, H \right)$. To facilitate the performance analysis, the single-antenna UEs are randomly located in the service area, following a uniform spatial distribution. The positions of the UE and the PA are denoted by $\boldsymbol{\psi} _u = \left( x_u, y_u, 0 \right)$ and $\boldsymbol{\psi} _p = \left( x_p, 0, H \right)$, respectively. The PA can be activated within the waveguide range, i.e., $x_p \in \mathcal{W} \triangleq \left[ 0, D_L \right]$. 

The distance between the PA and UE is given by
\begin{equation}\label{dist PA UE}
\left\| {{\boldsymbol{\psi} _p} - {\boldsymbol{\psi} _u}} \right\| = \sqrt {{\left( {x_p} - {x_u} \right)}^2 + d_0^2} \triangleq d,
\end{equation}
where ${d_0^2} = y_u^2 + {\left( H - {H_u} \right)}^2$, and $\left\| \cdot \right\|$ denotes the Euclidean distance.

\subsection{Blockage Model}
To facilitate the tractable performance analysis that incorporates both LoS blockage effect and NLoS propagation, a simplified scenario that includes obstacles with deterministic size and position is considered~\cite{xie2026pinchingantennasblockageawareenvironments, 10343131}. The set of obstacles is defined as $\mathcal{B} \triangleq \left\{ {\mathcal{B}_n} \right\}_{n=1}^{N_{\mathrm{obs}}}$, where $N_{\mathrm{obs}}$ denotes the total number of obstacles. To avoid analysis difficulties caused by differences in the shapes of obstacles, the obstacles are modeled as cylinders with heights of $H_n$. The projection of the $n$-th obstacle on the $x$–$y$ plane is defined as follows:
\begin{equation}\label{obs disk}
{\mathcal{B}_n} \triangleq \left\{ {\mathbf{p}} \in {\mathbb{R}^2} \,\middle|\, \left\| {\mathbf{p} - {\mathbf{b}_n}} \right\| \leqslant {r_n} \right\},
\end{equation}
where $\mathbf{b}_n={\left( {x_n,y_n,0} \right)}$ and $r_n$ denote the center and radius, respectively.

The occurrence of LoS blockage is strongly correlated with the positions of PAs, obstacles, and UEs. As illustrated in Fig.~\ref{fig: PASS under dynamic channels}(a), owing to the blockage effect of obstacles, part of the waveguide may fall into a ``LoS-blind'' region, in which PAs cannot establish LoS links with the UE. Such a region is defined as the blockage region induced by the $n$-th obstacle, which is denoted by $\mathcal{X}_n$.

A blockage region only forms when the obstacle is located between the waveguide and the UE. Given the positions of the UE and the $n$-th obstacle, the presence of blockage region can be determined by checking whether the inequalities $y_u y_n > 0$ and $\left| y_u \right| > \left| y_n \right| $ are satisfied simultaneously. When these conditions hold, the blockage region can be obtained by three steps: i) Determine two tangents from the UE to the cross-section of obstacle; ii) Calculate the intersection points of the two tangents with the waveguide line to obtain the initial endpoints, which are denoted by $x_n^{b - }$ and $x_n^{b + }$, respectively; and iii) Intersect the interval $\left[ x_n^{b - },x_n^{b + } \right]$ with the finite domain of the waveguide to yield the blockage region, i.e., $\mathcal{X}_n = \left[ x_n^{b - },x_n^{b + } \right] \cap \left[ 0,{D_L} \right]$.

For notational simplicity, when $\mathcal{X}_n \neq \varnothing$ holds, it is equivalently written as ${\mathcal{X}_n} = \left[ {x_n^{-},x_n^{+}} \right]$. If the LoS path is blocked by multiple obstacles, the total blockage region is the union of all blockage regions by individual obstacles, which is given by
\begin{equation}\label{total blockage region}
\mathcal{X} = \bigcup\limits_{n = 1}^{N_{\mathrm{obs}}} {\mathcal{X}_n},
\end{equation}

The LoS-feasible region is the complement of the total blockage region with respect to the waveguide range, i.e., $\mathcal{W} \backslash \mathcal{X}$. Accordingly, a binary discrete variable $\xi$ is defined as the LoS indicator, which is given by
\begin{equation}\label{LoS blockage condition}
\xi =
\begin{cases}
  0, & \mbox{if } x_p \in \mathcal{X}, \\
  1, & \mbox{otherwise}.
\end{cases}
\end{equation}

\subsection{Channel and Signal Model}
Depending on whether the LoS path between the PA and the UE is blocked, two channel conditions are investigated. On the one hand, as illustrated in Fig.~\ref{fig: PASS under dynamic channels}(b), when the LoS path exists, the propagation is dominated by a strong LoS component accompanied by scattered NLoS components, leading to a Rician fading model. On the other hand, as illustrated in Fig.~\ref{fig: PASS under dynamic channels}(a), when the LoS path is absent, only scattered multipath components remain, and the channel is modeled as a NLoS channel characterized by the Rayleigh fading.

\emph{1) LoS channel component:} From a PA to a UE, the free-space path loss governs the large-scale fading of the LoS channel. The small-scale fading is limited to the phase shift of signals propagating in free-space environments. The LoS channel component is given by
\begin{equation}\label{LoS component}
h_L = \frac{\sqrt \eta  }{d}{e^{ - j\frac{2\pi }{\lambda} d }},
\end{equation}
where $\eta = \tfrac{c^2}{\left(4\pi f_c \right)^2 }$, $c$ denotes the speed of light, $f_c$ denotes the carrier frequency, and $\lambda = \tfrac{c}{f_c}$ denotes the carrier wavelength.

\emph{2) NLoS channel component:} The NLoS component is modeled as the superposition of multiple independent scattering clusters, which is expressed as follows~\cite{xu2025pinchingantennadesignrandomlos}:
\begin{equation}\label{NLoS component}
h_N = \sum\limits_{k = 1}^K {g_{N,k}},
\end{equation}
where $K$ denotes the number of clusters, and $g_{N,k}$ denotes the small-scale fading gain of the $k$-th cluster. Although each NLoS path involves at least two-hop propagation via scatterers, the cluster-based representation provides an accurate statistical approximation of multipath propagation. Each cluster is modeled as an independent zero-mean complex Gaussian variable, which is expressed as follows:
\begin{equation}\label{NLoS small-scale fading k}
g_{N,k} \sim \mathcal{CN}\left( {0,\frac{\mu _{N,k}^2}{d^{\alpha _N}}} \right),
\end{equation}
where $\mu _{N,k}^2$ denotes the normalized average power of the $k$-th cluster, and $\alpha _N$ denotes the path loss exponent of NLoS paths. Accordingly, the average NLoS channel power can be expressed as follows:
\begin{equation}\label{NLoS power}
\mathbb{E}\left[ {\left| {h_N} \right|}^2 \right] = \frac{1}{d^{\alpha _N}}\sum\limits_{k = 1}^K {\mu _{N,k}^2} \triangleq \frac{\mu _N^2}{d^{\alpha _N}},
\end{equation}
where $\mu _N^2 \triangleq \sum\nolimits_{k = 1}^K {\mu _{N,k}^2}$ denotes the effective NLoS cluster powers, or the average NLoS power. Thus, the NLoS channel component also follows a zero-mean complex Gaussian distribution, which is expressed as follows:
\begin{equation}\label{NLoS component Gaussian}
{h_N} \sim \mathcal{CN}\left( 0, \sigma_N^2 \right),
\end{equation}
where $\sigma_N^2 = \mu _N^2 d^{-\alpha _N}$.

\emph{3) In-waveguide channel component:} As the signal propagates along the waveguide, it experiences an exponential power attenuation as well as a distance-dependent phase shift~\cite{xu2025pinchingantennasystemsinwaveguideattenuation}. Therefore, the in-waveguide channel gain is given by
\begin{equation}\label{in-waveguide channel gain}
h_g = e^{ - \left( {\alpha _g} + j\frac{2\pi }{\lambda _g} \right)\left\| {\boldsymbol{\psi} _0} - {\boldsymbol{\psi} _p} \right\|},
\end{equation}
where $\alpha_g$ denotes the in-waveguide attenuation coefficient, and $\lambda _g = \tfrac{\lambda}{n_{\mathrm{eff}}}$ denotes the in-waveguide carrier wavelength with $n_{\mathrm{eff}}$ being the effective refractive index of a dielectric waveguide.

\emph{4) Composite channel and signal model:} In the presence of both in-waveguide attenuation and LoS blockage, the effective channel gain consists of the in-waveguide and spatial channel gains, which can be expressed as follows:
\begin{equation}\label{effective channel gain}
h = {h_g}\left( \xi {h_L} + {h_N} \right).
\end{equation}

Note that when $\xi = 0$, the LoS component is absent, where the spatial channel gain is characterized by a NLoS channel with Rayleigh distribution. Conversely, when $\xi = 1$, the spatial channel gain is characterized by a LoS-dominated channel with Rician distribution.

The Rician factor can be written as follows:
\begin{equation}\label{Rician K-factor}
\kappa = \frac{\xi {\left| h_L \right|}^2 }{\mathbb{E}\left[ {\left| h_N \right|}^2 \right]} = \frac{\xi \eta {d^{{\alpha _N} - 2}}}{\mu _N^2},
\end{equation}
which is defined by the expected power ratio between LoS and NLoS components. Here, the Rician factor depends on multiple factors including the presence of LoS blockage, the average NLoS power, and the PA-to-UE distance.

The signal received at the UE is expressed as follows:
\begin{equation}\label{received signal at the UE}
r = h\sqrt P s + n,
\end{equation}
where $P$ denotes the transmit power budget, $s$ represents the transmit signal of the UE with $\mathbb{E} [ {\left| s \right|}^2 ] = 1$, and $n \sim \mathcal{CN} \left(0,\sigma_n^2 \right)$ denotes the additive white Gaussian noise (AWGN) with power $\sigma_n^2$.

\section{Performance Analysis}
This section provides an analytical study on the performance of PASS under dynamic channels with blockages. The following three subsections present the analysis of outage probabilities, ergodic rates, and optimal PA positions.

\subsection{Outage Probability}

The outage probability is defined as the probability that the received SNR at the UE is lower than the required threshold, which is given by
\begin{equation}\label{OP}
{\mathbb{P}^{\mathrm{out}}} = \mathbb{P}\left( {\gamma  \leqslant {\gamma _{\mathrm{th}}}} \right),
\end{equation}
where $\gamma = \tfrac{P {\left| h \right|}^2}{\sigma _n^2}$ and $\gamma _{\mathrm{th}}$ denote the received SNR and its threshold at the UE, respectively.

When the LoS path is blocked, i.e., $\xi = 0$, the channel gain reduces to $h = {h_g}{h_N}$. Thus, the outage probability of the UE can be derived in the following theorem.
\begin{theorem}\label{theorem: OP Rayleigh}
Consider a single-PA single-UE PASS that operates over a Rayleigh fading channel. For deterministic positions of the PA and UE, the outage probability of the UE can be expressed as follows:
\begin{equation}\label{OP Rayleigh}
\mathbb{P}_N^{\mathrm{out}} = 1 - \exp \left( - \frac{\gamma _{\mathrm{th}}{d^{\alpha _N}}}{\rho \mu _N^2} \right),
\end{equation}
where $\rho  = \tfrac{P}{\sigma _n^2}{{\left| h_g \right|}^2}$ denotes the transmit SNR at the PA.
\begin{proof}
For the Rayleigh fading channel, the outage probability of the UE is given by
\begin{equation}\label{OP Rayleigh 1}
\mathbb{P}_N^{\mathrm{out}} = \mathbb{P}\left( {{\left| h_N \right|}^2 \leqslant \frac{\gamma _{\mathrm{th}}} \rho} \right).
\end{equation}

According to the cumulative distribution function (CDF) of Rayleigh distribution, ${\left| h_N \right|}^2$ follows an exponential distribution, which can be given by
\begin{equation}\label{CDF Rayleigh channel power}
F_{{\left| {h_N} \right|}^2} \left( z \right) = 1 - \exp \left( { - \frac{z}{\sigma _N^2}} \right),\ z > 0,
\end{equation}
and the probability density function (PDF) of ${\left| h_N \right|}^2$ is given by
\begin{equation}\label{PDF Rayleigh channel power}
f_{{\left| {h_N} \right|}^2} \left( z \right) = \frac{1}{\sigma _N^2}\exp \left( { - \frac{z}{\sigma _N^2}} \right),\ z > 0.
\end{equation}

Note that the outage probability can be equivalently expressed by using the CDF in~\eqref{CDF Rayleigh channel power}, which is given by
\begin{equation}\label{OP Rayleigh 2}
\mathbb{P}_N^{\mathrm{out}} = F_{{\left| {h_N} \right|}^2} \left( \frac{\gamma _{\mathrm{th}}}{\rho } \right).
\end{equation}
Hence, the results in~\eqref{OP Rayleigh} is readily obtained, and the proof is complete.
\end{proof}
\end{theorem}

For the Rician fading channel, the outage probability of the UE can be derived in a closed-form expression, as summarized in the following theorem.
\begin{theorem}\label{theorem: OP Rician}
Consider a single-PA single-UE PASS that operates over a Rician fading channel. For deterministic positions of the PA and UE, the outage probability of the UE can be expressed as follows:
\begin{equation}\label{OP Rician}
\begin{split}
\mathbb{P}_L^{\mathrm{out}} = 1 - {Q_1}\left( {\frac{\sqrt {2\eta {d^{\alpha _N - 2}}} }{\mu _N},\frac{1}{\mu _N}\sqrt {\frac{2{\gamma _{\mathrm{th}}}{d^{\alpha _N}}}{\rho }} } \right),
\end{split}
\end{equation}
where ${Q_1}\left(  \cdot  \right)$ denotes the first-order Marcum-Q function~\cite{simon2002probability}.
\begin{proof}
For the Rician fading channel, the outage probability can be written as follows:
\begin{equation}\label{OP Rician 1}
\mathbb{P}_L^{\mathrm{out}} = \mathbb{P}\left( {\left| h_L + h_N \right|}^2 \leqslant \frac{\gamma _{\mathrm{th}}}\rho \right).
\end{equation}

According to the CDF of Rician distribution, ${\left| h_L + h_N \right|}^2$ follows a non-central Chi-square distribution, which can be given by
\begin{equation}\label{CDF Rician channel power}
F_{{\left| {h_L} + {h_N} \right|}^2}\left( z \right) = 1 - {Q_1}\left( {\sqrt {\frac{2{\left| {{h_L}} \right|}^2}{\sigma _N^2}} ,\sqrt {\frac{2z}{\sigma _N^2}} } \right),\  z > 0,
\end{equation}
and the PDF of ${\left| h_L + h_N \right|}^2$ is given by
\begin{equation}\label{PDF Rician channel power}
\begin{split}
&{f_{{{\left| {{h_L} + {h_N}} \right|}^2}}} \left( z \right) \\
     & = \frac{1}{{\sigma _N^2}}\exp \left( { - \frac{{z + {{\left| {{h_L}} \right|}^2}}}{{\sigma _N^2}}} \right){I_0}\left( {\frac{{2\sqrt {{{\left| {{h_L}} \right|}^2}z} }}{{\sigma _N^2}}} \right),\ z > 0,
\end{split}
\end{equation}
where $I_0\left( \cdot \right)$ denotes the zero-order modified Bessel function of the first kind~\cite{Table_of_integrals}.

Note that the outage probability can be equivalently expressed by using the CDF in~\eqref{CDF Rician channel power}, which is given by
\begin{equation}\label{OP Rician 2}
\mathbb{P}_L^{\mathrm{out}} = F_{{\left| {{h_L} + {h_N}} \right|}^2}\left( {\frac{\gamma _{\mathrm{th}}}{\rho }} \right).
\end{equation}
Hence, the results in~\eqref{OP Rician} is readily obtained, and the proof is complete.
\end{proof}
\end{theorem}

\subsection{Ergodic Rate}

The ergodic rate explicitly characterizes the performance averaged over the fading randomness. The ergodic rate of the UE is given by
\begin{equation}\label{Rate original}
R = {\mathbb{E}_h}\left[ {{\log }_2}\left( 1 + \frac{P}{\sigma _n^2}{\left| h \right|}^2 \right) \right].
\end{equation}

When the LoS component is absent, the received signal is supported by NLoS propagation. A closed-form expression of the ergodic rate can be derived, which is stated in the following theorem.
\begin{theorem}\label{theorem: ER Rayleigh}
Consider a single-PA single-UE PASS that operates over a Rayleigh fading channel. For deterministic positions of the PA and UE, the ergodic rate of the UE can be expressed as follows:
\begin{equation}\label{ER Rayleigh}
R_N = \frac{1}{\ln 2}\exp \left( \frac{d^{\alpha _N}}{\rho \mu _N^2} \right){E_1}\left( \frac{d^{\alpha _N}}{\rho \mu _N^2} \right),
\end{equation}
where $E_1 \left( \cdot \right)$ denotes the exponential integral function, defined as ${E_1}\left( x \right) \triangleq \int_x^\infty  {\tfrac{e^{ - t}}{t}dt}, x > 0$.
\begin{proof}
For the Rayleigh fading channel, the ergodic rate is given by
\begin{equation}\label{ER Rayleigh 1}
\begin{split}
R_N & = \mathbb{E}_{{\left| h_N \right|}^2} \left[ {\log }_2 \left( 1 + \rho {\left| {h_N} \right|}^2 \right) \right] \\
    & = \int_0^\infty  {f_{{\left| h_N \right|}^2} \left( z \right){{\log }_2}\left( {1 + \rho z} \right)dz}.
\end{split}
\end{equation}

Substituting~\eqref{PDF Rayleigh channel power} into~\eqref{ER Rayleigh 1} yields~\eqref{ER Rayleigh}. Hence, the proof is complete.
\end{proof}
\end{theorem}

For given PA and UE positions, it is worth emphasizing that both the LoS component and the in-waveguide attenuation are deterministic. The randomness of effective channel gain solely originates from the NLoS component, whose mean and variance are deterministic. Thus, under a Rician fading channel, the ergodic rate of the UE can be derived in the following theorem.
\begin{theorem}\label{theorem: ER Rician}
Consider a single-PA single-UE PASS that operates over a Rician fading channel. For deterministic positions of the PA and UE, the ergodic rate of the UE can be expressed as follows:
\begin{equation}\label{ER Rician}
R_L = \frac{\rho }{\ln 2}\int_0^\infty  {\frac{1}{1 + \rho z}{Q_1}\left( {\frac{\sqrt {2\eta {d^{{\alpha _N} - 2}}} }{{\mu _N}},\frac{\sqrt {2{d^{\alpha _N}}z} }{{\mu _N}}} \right)dz},
\end{equation}
\begin{proof}
For the Rician fading channel, the ergodic rate is given by
\begin{equation}\label{ER Rician 1}
\begin{split}
R_L & = \mathbb{E}_{{\left| h_L + h_N \right|}^2} \left[ {\log }_2 \left( 1 + \rho {\left| {h_L + h_N} \right|}^2 \right) \right] \\
    & = \frac{\rho }{\ln 2}\int_0^\infty  {\frac{1 - {F_{{\left| {{h_L} + {h_N}} \right|}^2}}\left( z \right)}{1 + \rho z}dz}.
\end{split}
\end{equation}

Substituting~\eqref{CDF Rician channel power} into~\eqref{ER Rician 1} yields~\eqref{ER Rician}. Hence, the proof is complete.
\end{proof}
\end{theorem}

Since it is challenging to obtain a closed-form expression of the ergodic rate in~\eqref{ER Rician}, a Taylor expansion~\cite{Table_of_integrals} is used to obtain a closed-form approximation in the following corollary.
\begin{corollary}\label{coro: ER Rician approximate}
Consider a single-PA single-UE PASS that operates over a Rician fading channel. For deterministic positions of the PA and UE, the ergodic rate of the UE can be approximated as follows:
\begin{equation}\label{ER Rician approximate}
\begin{split}
& R_L \approx  {\log _2}\left( {1 + \rho \left( {\frac{\eta }{d^2} + \frac{\mu _N^2}{d^{\alpha _N}}} \right)} \right) \\
     & - \frac{{{\rho ^2}\mu _N^2}}{{2{d^{{\alpha _N}}}{{\left( {1 + \rho \left( {\frac{\eta }{{{d^2}}} + \frac{{\mu _N^2}}{{{d^{{\alpha _N}}}}}} \right)} \right)}^2}\ln 2}}\left( {\frac{{2\eta }}{{{d^2}}} + \frac{{\mu _N^2}}{{{d^{{\alpha _N}}}}}} \right).
\end{split}
\end{equation}
\begin{proof}
Please refer to Appendix A.
\end{proof}
\end{corollary}

\begin{remark}\label{remark: NLoS twofold effect}
The effect of NLoS propagation on the performance of PASS is twofold and critically depends on the relative strength between the LoS and NLoS links, which can be characterized by the Rician factor. In the LoS-dominated regimes, NLoS scattering primarily introduces randomness and makes a minimal contribution to the performance. However, when the LoS component is weak or blocked, NLoS propagation plays a constructive role in maintaining the communication link and preventing outages.
\end{remark}

\begin{proposition}\label{proposition: Rician Rate greater than LoS Rate}
For a fixed transmit power and path-loss geometry, by accounting for both the LoS and NLoS components, the ergodic rate achieved under a Rician fading model is strictly greater than that achieved under a LoS propagation-based evaluation.
\begin{proof}
Please refer to Appendix B.
\end{proof}
\end{proposition}

\begin{remark}\label{remark: if NLoS can be ignored}
Although NLoS scatterers cause signal power fluctuations and random phase shifts, \textbf{Proposition~\ref{proposition: Rician Rate greater than LoS Rate}} reveals that NLoS propagation can enhance the average performance in single-PA single-user scenarios. For simplicity, it is reasonable to ignore the gain fluctuations caused by NLoS scattering if the LoS component dominates the channel gain. However, ignoring NLoS propagation, particularly when the LoS component is weak (i.e., when the Rician factor is small), leads to an underestimation of PASS performance.
\end{remark}

\subsection{Optimal PA Position in a Blockage Scenario}

When the PA is activated at its optimal position, it yields the maximum possible ergodic rate for the UE. The optimal PA position is given by
\begin{equation}\label{problem of optimal PA position}
x_p^{\mathrm{opt}} = \displaystyle \arg \max \limits_{{x_p} \in \mathcal{W}} R.
\end{equation}

It is quite challenging to derive a closed-form expression for the optimal PA position in~\eqref{problem of optimal PA position}. Specifically, the challenge stems not only from comparing the achievable performance under Rayleigh and Rician fading channel conditions, but also from the intrinsic tradeoff between spatial propagation loss and in-waveguide attenuation. To tackle the challenge on deriving the optimal PA position, a three-step analytical procedure is adopted: i) Identify the blockage region within the service area; ii) Derive the candidate optimal PA positions inside and outside the blockage region, respectively; and iii) Among the candidate optimal PA positions, compare their achievable performance to determine the final optimal PA position. The above steps are applicable in both continuous and discrete placement of PA positions on the waveguide.

The following lemma yields expressions for the two endpoints of the blockage region.
\begin{lemma}\label{lemma: the intersections pf blockage region}
Consider the blockage region on the waveguide induced by the $n$-th obstacle, which is regarded as a continuous segment defined by the closed interval $[x_n^-, x_n^+]$. The two endpoints of the blockage region are expressed as follows:
\begin{equation}\label{the intersections pf blockage region}
x_n^ -  = \max \left\{ {0,x_n^{b - }} \right\},x_n^ +  = \min \left\{ {{D_L},x_n^{b + }} \right\},
\end{equation}
where $x_n^{b - }$ is given by
\begin{equation}\label{initial intersection 1 lemma}
x_n^{b - }  =
\begin{cases}
\min \left\{ {x_u} , {x_u} - \frac{y_u}{m_3} \right\}, & \mbox{if } \Delta {x_n} = {r_n}, \\
\min \left\{ {x_u} - \frac{y_u}{m_1}, {x_u} - \frac{y_u}{m_2} \right\}, & \mbox{otherwise},
\end{cases}
\end{equation}
$x_n^{b + }$ is given by
\begin{equation}\label{initial intersection 2 lemma}
x_n^{b + }  =
\begin{cases}
\max \left\{ {x_u} , {x_u} - \frac{y_u}{m_3} \right\}, & \mbox{if } \Delta {x_n} = {r_n}, \\
\max \left\{ {x_u} - \frac{y_u}{m_1}, {x_u} - \frac{y_u}{m_2} \right\}, & \mbox{otherwise},
\end{cases}
\end{equation}
$m_1$ is given by
\begin{equation}\label{tangents' slope general 1 lemma}
m_1 = \frac{{\Delta {x_n}\Delta {y_n} + {r_n}\sqrt {{{\left( {\Delta {x_n}} \right)}^2} + {{\left( {\Delta {y_n}} \right)}^2} - r_n^2} }}{{{\left( {\Delta {x_n}} \right)}^2} - r_n^2},
\end{equation}
$m_2$ is given by
\begin{equation}\label{tangents' slope general 2 lemma}
m_2 = \frac{{\Delta {x_n}\Delta {y_n} - {r_n}\sqrt {{{\left( {\Delta {x_n}} \right)}^2} + {{\left( {\Delta {y_n}} \right)}^2} - r_n^2} }}{{{\left( {\Delta {x_n}} \right)}^2} - r_n^2},
\end{equation}
$m_3$ is given by
\begin{equation}\label{tangents' slope special lemma}
m_3 = \frac{{{\left( {\Delta {y_n}} \right)}^2} - r_n^2}{2 r_n \Delta {y_n}},
\end{equation}
$\Delta {x_n} = {x_n} - {x_u}$, and $\Delta {y_n} = {y_n} - {y_u}$.
\begin{proof}
Please refer to Appendix C.
\end{proof}
\end{lemma}

\emph{1) The candidate optimal PA position under a Rayleigh fading channel:} Within a blockage region, the average received SNR at the UE can be regarded as a function of the PA position, which is expressed as follows:
\begin{equation}\label{average SNR Rayleigh}
{\gamma _N}\left( {x_p} \right) = \rho \mathbb{E}\left[ {{{\left| {{h_N}} \right|}^2}} \right] = \frac{P \mu _N^2 {e^{ - 2{\alpha _g}{x_p}}}}{\sigma _n^2{d^{{\alpha _N}}}\left( {{x_p}} \right)}.
\end{equation}

Since the ergodic rate in~\eqref{ER Rayleigh} is monotonically increasing with the average received SNR, maximizing the ergodic rate is equivalent to maximizing the average received SNR. When the transmit power of the AP and the average NLoS power are fixed, maximizing the average received SNR is equivalent to maximizing the total channel gain as well. Therefore, the problem of solving the optimal PA position can be simplified to maximizing total channel gain, which is expressed as follows:
\begin{equation}\label{problem of optimal PA position Rayleigh}
\begin{split}
\arg \mathop {\max }\limits_{{x_p} \in {\mathcal{X}_n}} {R_N} & = \arg \mathop {\max }\limits_{{x_p} \in {\mathcal{X}_n}} {\gamma _N}\left( {x_p} \right)  \\
     & = \arg \mathop {\max }\limits_{{x_p} \in {\mathcal{X}_n}} \frac{e^{ - 2{\alpha _g}{x_p}}}{d^{\alpha _N}\left( x_p \right)}.
\end{split} 
\end{equation}

Solving the problem in~\eqref{problem of optimal PA position Rayleigh} yields closed-form expressions for the optimal PA position, as summarized in the following proposition.
\begin{proposition}\label{proposition: candidate optimal pos of PA Rayleigh}
Within a blockage region induced by the $n$-th obstacle, i.e., under a Rayleigh fading channel, to balance spatial propagation loss and in-waveguide attenuation, the optimal PA position along the waveguide is obtained by maximizing total channel gain, which is expressed as follows:
\begin{equation}\label{candidate optimal pos of PA Rayleigh}
x_p^{\mathrm{opt}} =
\begin{cases}
  x_{p2}, & \mbox{if } \alpha _N^2 > 4\alpha _g^2d_0^2 \mbox{ and } x_n^- \leqslant x_{p2} \leqslant x_n^+, \\
  x_{p3}, & \mbox{if } \alpha _N^2 > 4\alpha _g^2d_0^2 \mbox{ and } x_n^+ < x_{p2}, \\
  x_n^-, & \mbox{otherwise,}
\end{cases}
\end{equation}
where $x_{p2} = {x_u} - \frac{{\alpha _N} - \sqrt {\alpha _N^2 - 16\alpha _g^2d_0^2} }{4{\alpha _g}}$, and $x_{p3} = \displaystyle \arg \max \limits_{x_p \in \left\{ {x_n^-, x_n^+} \right\}} d^{-\alpha _N} e^{ - 2{\alpha _g}{x_p}}$.
\begin{proof}
Please refer to Appendix D.
\end{proof}
\end{proposition}

\begin{remark}\label{remark: the optimal PA position mathemetical rule Rayleigh}
In addition to the constraints imposed by the blockage region, the optimal PA position is influenced by multiple factors, including the path loss coefficient, the loss coefficient within the waveguide, and the relative position of the user in relation to the waveguide.
\end{remark}

\emph{2) The candidate optimal PA position under a Rician fading channel:} Since the ergodic rate in~\eqref{ER Rician} is not in closed form, it is challenging to derive a closed-form expression for the optimal PA position in a LoS-feasible region. Due to the concavity of the logarithmic function, the following inequality holds true by using Jensen's inequality~\cite{Table_of_integrals}:
\begin{equation}\label{ER Jensen to average SNR Rician}
\mathbb{E}\left[ {{{\log }_2}\left( 1 + \rho {\left| h \right|}^2 \right)} \right] \leqslant {\log _2}\left( 1 + \rho \mathbb{E}\left[ {\left| h \right|}^2 \right] \right).
\end{equation}

In the LoS-dominated regime, although the NLoS component remains stochastic, its contribution to the channel gain becomes marginal compared to that of the deterministic LoS component. As a result, the instantaneous channel power experiences only minor fluctuations around its average value, and it can be well approximated by its mean, i.e., $\left| h \right|^2 \to \mathbb{E}[ \left| h \right|^2 ]$. Then, the data rate that is obtained based on the average received SNR closely approximates the ergodic rate, which indicates that the two sides of the inequality in~\eqref{ER Jensen to average SNR Rician} become nearly equal.

Under a Rician fading channel, the average received SNR at the UE can be regarded as a function of the PA position, which is expressed as follows:
\begin{equation}\label{average SNR Rician}
\begin{split}
{\gamma _L}\left( {x_p} \right) & = \rho \mathbb{E}\left[ {\left| {{h_L} + {h_N}} \right|}^2 \right] \\
     & = \frac{P{e^{ - 2{\alpha _g}{x_p}}}}{\sigma _n^2}\left( \frac{\eta }{d^2 \left( x_p \right)} + \frac{\mu _N^2}{d^{\alpha _N} \left( x_p \right)} \right).
\end{split}
\end{equation}

When the transmit power and the average NLoS power are fixed, maximizing the average received SNR is equivalent to maximizing the total channel gain. Therefore, the problem of solving optimal PA position can be approximately simplified to maximizing total channel gain, which is expressed as follows:
\begin{equation}\label{problem of optimal PA position Rician}
\begin{split}
\arg & \mathop {\max }\limits_{{x_p} \in \mathcal{W}\backslash \mathcal{X}} {R_L} \approx \arg \mathop {\max }\limits_{x_p \in \mathcal{W}\backslash \mathcal{X}} {\gamma _L}\left( {x_p} \right) \\
     & = \arg \mathop {\max }\limits_{{x_p} \in \mathcal{W}\backslash \mathcal{X}} {e^{ - 2{\alpha _g}{x_p}}}\left( {\frac{\eta }{{d^2}\left( x_p \right)} + \frac{\mu _N^2}{d^{\alpha _N}\left( {x_p} \right)}} \right).
\end{split}
\end{equation}

By solving the problem in~\eqref{problem of optimal PA position Rician} via a simple one-dimensional linear search, the candidate optimal PA position in the LoS-feasible region can be efficiently obtained.

\begin{table}[t]
  \begin{center}
  \caption{Simulation Parameters}\footnotesize\label{tab: simu para}
  \begin{tabular}{|l|l|l|}
    \hline
    \textbf{Parameter} & \textbf{Description} & \textbf{Value} \\
    \hline
    $f_c$ & Carrier frequency & $28$ GHz \\    
    \hline
    $BW$ & Bandwidth & $10$ MHz \\
    \hline
    $P$ & Transmit power & $10$ dBm \\
    \hline
    $\sigma _n^2$ & AWGN power & $-104$ dBm \\
    \hline
    $n_{\mathrm{eff}}$ & Effective refractive & $1.4$ \\
    & index of the waveguide & \\
    \hline
    $H$ & Height of the waveguide & $3$ m \\    
    \hline
    $D_L$ & Length of the waveguide/service area & $10$ m \\
    \hline
    $D_W$ & Width of the service area & $10$ m \\
    \hline
    $\gamma _{\mathrm{th}}$ & Received SNR threshold & $0$ dB \\    
    \hline
    $\mu _N^2$ & Average NLoS power & $-50$ dBm \\
    \hline
    $\alpha _N$ & Path loss exponent of NLoS & $3$ \\
    \hline
    $\alpha_g$ & In-waveguide attenuation & $1.47$ dB/m \\
    & coefficient & \\
    \hline
  \end{tabular}
  \end{center}
\end{table}

\begin{figure*}[t!]
    \centering
    \begin{minipage}{0.32\textwidth}
        \centering
        \includegraphics[width = \linewidth]{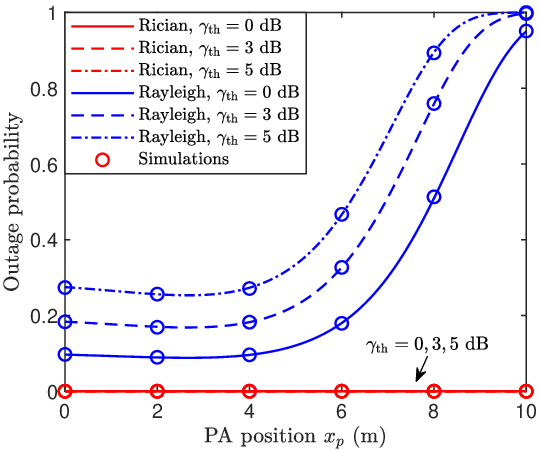}        
        \centerline{\footnotesize(a)}
    \end{minipage}
    \hfill
    \begin{minipage}{0.32\textwidth}
        \centering
        \includegraphics[width = \linewidth]{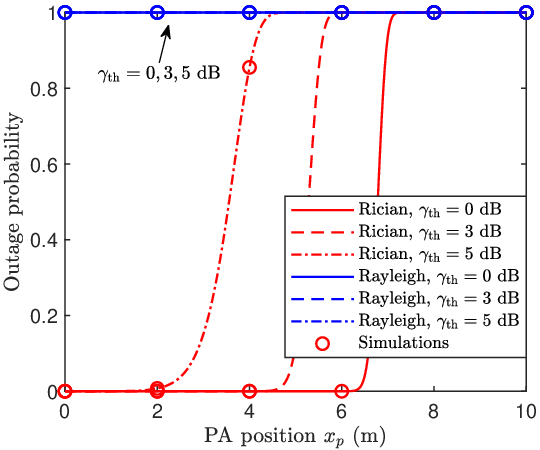}        
        \centerline{\footnotesize(b)}
    \end{minipage}
    \hfill
    \begin{minipage}{0.32\textwidth}
        \centering
        \includegraphics[width = \linewidth]{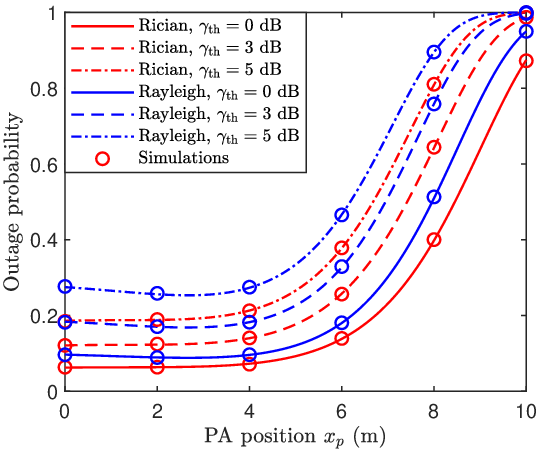}        
        \centerline{\footnotesize(c)}
    \end{minipage}
    \caption{Outage probability versus PA position in different SNR regimes: (a) in the high-SNR regime with the transmit power of $10$ dBm and the average NLoS power of $-50$ dBm; (b) in the low-SNR regime with the transmit power of $-20$ dBm and the average NLoS power of $-50$ dBm; and (c) in the low-SNR regime with the transmit power of $-20$ dBm and the average NLoS power of $-20$ dBm.}
    \label{fig: OP versus PA pos comparing SNR_th}
\end{figure*}

\begin{figure}[t!]
    \centering
    \includegraphics[width = 0.32\textwidth]{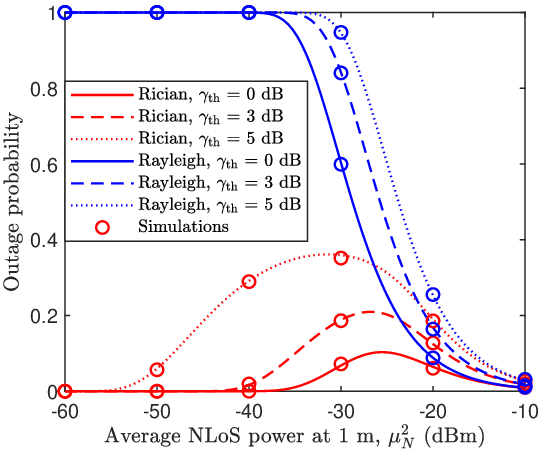}
    \caption{Sensitivity of the outage probability to the average NLoS power, where the transmit power is set to $-20$ dBm.}
    \label{fig: OP versus NLoS power comparing SNR_th}
\end{figure}

\section{Numerical Results}

This section provides numerical evaluations of the performance of PASS under dynamic channels, where analytical results are validated against Monte Carlo simulations. The in-waveguide attenuation corresponds to approximately $1.47$ dB/m at $28$ GHz when polytetrafluoroethylene is used as the dielectric waveguide~\cite{pozar2021microwave}. The AWGN power is configured by using $\sigma _n^2 = - 174 + 10{\log _{10}}\left( {BW} \right)$, which is related to the bandwidth in practice. Unless otherwise specified, the parameters in Table~\ref{tab: simu para} are used~\cite{ding2025flexible}.

\subsection{Outage Probability Under Rayleigh and Rician Fading Channels}

Fig.~\ref{fig: OP versus PA pos comparing SNR_th} illustrates the outage probability as a function of the PA position under different channel models in the high-SNR and low-SNR regime. The UE is located at $\left( 5, 2.5, 0 \right)$ m. The Monte Carlo simulations closely match the analytical results, which verifies the accuracy of \textbf{Theorem~\ref{theorem: OP Rayleigh}} and \textbf{Theorem~\ref{theorem: OP Rician}}. In Fig.~\ref{fig: OP versus PA pos comparing SNR_th}(a), where PASS operates in the high-SNR regime and with a large Rician factor, the outage probability under the Rician fading channel approaches 0 for different received SNR thresholds, which highlights the robustness of the LoS component at sufficiently high transmit power. Under the Rayleigh fading channel, the outage probability changes significantly as the received SNR threshold increases. In Fig.~\ref{fig: OP versus PA pos comparing SNR_th}(b), when PASS operates in the low-SNR regime but still with a large Rician factor, the outage probability under the Rician fading channel significantly increases when the PA's $x$-position is greater, which indicates that a low transmit power limits reliable communication within a small PA region near the AP. Meanwhile, the outage probability under the Rayleigh fading channel approaches 1 for different received SNR thresholds, which indicates that even in the presence of strong NLoS scattering with rich multipath components, the resulting channel fluctuations can still cause severe fading. Therefore, the presence of a LoS component is crucial for stabilizing the channel gain and improving the outage performance, regardless of the transmit power or the activated PA position. In Fig.~\ref{fig: OP versus PA pos comparing SNR_th}(c), when the transmit power is $-20$ dBm but the average NLoS power is $-20$ dBm, the outage probability increases monotonically with the PA position, yet the transition is much smoother than that in Fig.~\ref{fig: OP versus PA pos comparing SNR_th}(b), since the enhanced scattered NLoS component introduces a more gradual variation in the effective channel gain. Overall, these results demonstrate that the PA position has a significant impact on outage performance, especially in the low-SNR regime and in channels without a strong LoS component.

Fig.~\ref{fig: OP versus NLoS power comparing SNR_th} illustrates the outage probability versus the average NLoS power for different outage thresholds under Rayleigh and Rician fading channels. The PA is located at $\left( 2.66, 0, 3 \right)$ m. A fundamental difference can be seen between the Rayleigh and Rician cases. On the one hand, under the Rayleigh fading channel, the outage probability decreases monotonically as the average NLoS power increases, since increasing the strength of NLoS scattering directly improves the received SNR statistics. As a result, the probability that the instantaneous SNR falls below the SNR threshold is gradually reduced from nearly 1 to nearly 0. On the other hand, under the Rician fading channel, the outage probability exhibits a single peak, namely, it initially increases from a value close to 0, then reaches a maximum, and finally decreases toward 0. Specifically, when the average NLoS power is very small (i.e., $-60$ dBm), the channel gain is dominated by the LoS component, and the received signal becomes relatively stable. As the average NLoS power increases, the NLoS scattering becomes pronounced, which introduces random phase shifts and strong fading fluctuations to the signal. As a result, the received SNR is more likely to fall below the SNR threshold. As the average NLoS power further increases (i.e., more than $-20$ dBm), the received SNR is sufficiently enhanced such that the probability of falling below the SNR threshold decreases. Overall, under the Rician fading channel, the outage probability demonstrates the interplay between average power enhancement and random fading intensification, which is consistent with \textbf{Remark~\ref{remark: NLoS twofold effect}}.

\begin{figure*}[t!]
    \centering
    \begin{minipage}{0.32\textwidth}
        \centering
        \includegraphics[width = \linewidth]{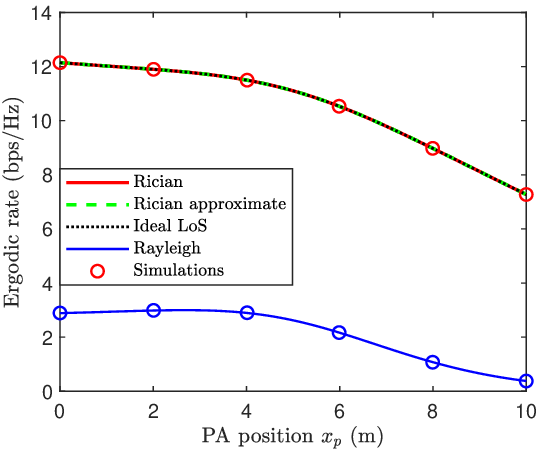}        
        \centerline{\footnotesize(a)}
    \end{minipage}
    \hfill
    \begin{minipage}{0.32\textwidth}
        \centering
        \includegraphics[width = \linewidth]{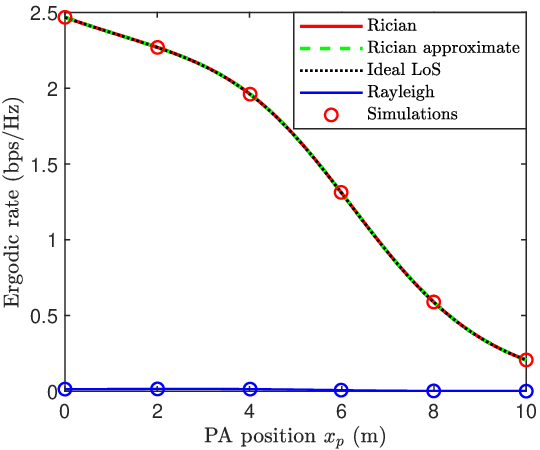}        
        \centerline{\footnotesize(b)}
    \end{minipage}
    \hfill
    \begin{minipage}{0.32\textwidth}
        \centering
        \includegraphics[width = \linewidth]{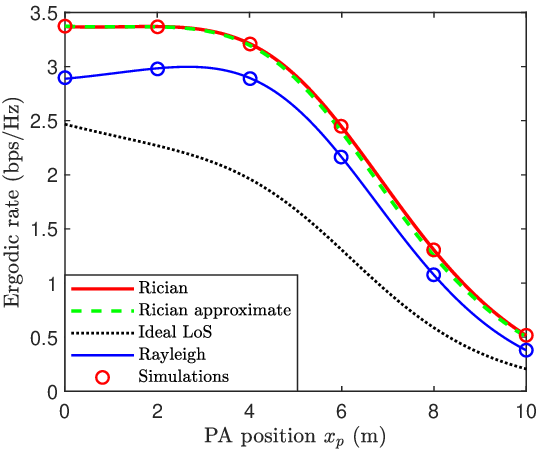}        
        \centerline{\footnotesize(c)}
    \end{minipage}
    \caption{Ergodic rate versus PA position in different SNR regimes: (a) in the high-SNR regime with the transmit power of $10$ dBm and the average NLoS power of $-50$ dBm; (b) in the low-SNR regime with the transmit power of $-20$ dBm and the average NLoS power of $-50$ dBm; and (c) in the low-SNR regime with the transmit power of $-20$ dBm and the average NLoS power of $-20$ dBm.}
    \label{fig: ER versus PA pos}
\end{figure*}

\begin{figure*}[t!]
    \centering
    \begin{minipage}{0.32\textwidth}
        \centering
        \includegraphics[width = \linewidth]{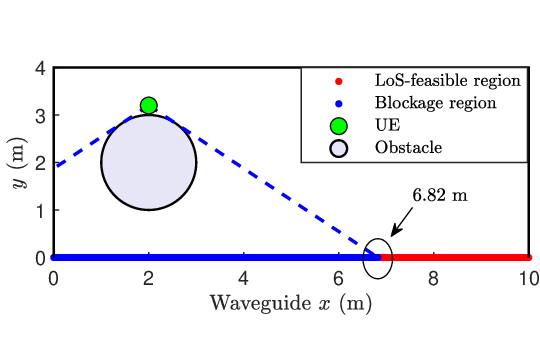}        
        \centerline{\footnotesize(a)}
    \end{minipage}
    \hfill
    \begin{minipage}{0.32\textwidth}
        \centering
        \includegraphics[width = \linewidth]{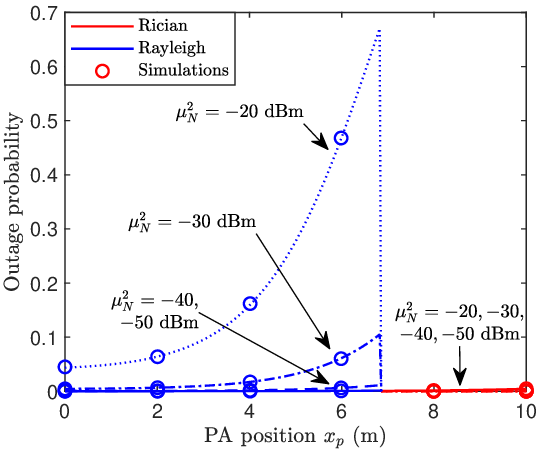}        
        \centerline{\footnotesize(b)}
    \end{minipage}
    \hfill
    \begin{minipage}{0.32\textwidth}
        \centering
        \includegraphics[width = \linewidth]{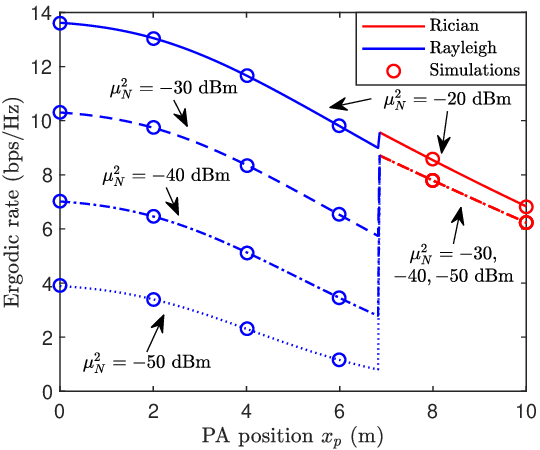}        
        \centerline{\footnotesize(c)}
    \end{minipage}   
    \caption{Illustrations of PASS in a blockage scenario and the corresponding performance: (a) The blockage region and the LoS-feasible region along the waveguide; (b) Outage probability versus PA position; and (c) Ergodic rate versus PA position.}
    \label{fig: nearPA_or_farPA scenario}
\end{figure*}

\subsection{Ergodic Rate Under Rayleigh and Rician Fading Channels}

Fig.~\ref{fig: ER versus PA pos} illustrates the ergodic rate versus the PA position under different SNR regimes for Rician and Rayleigh fading channels. The Monte Carlo simulations closely match the analytical results, which verifies the accuracy of \textbf{Theorem~\ref{theorem: ER Rayleigh}} and \textbf{Theorem~\ref{theorem: ER Rician}}. It can be also observed that the analytical results and the approximate results under Rician fading channels are almost indistinguishable, which verifies the high accuracy of~\textbf{Corollary~\ref{coro: ER Rician approximate}}.

In Fig.~\ref{fig: ER versus PA pos}(a), where PASS operates in the high-SNR regime and with a large Rician factor, the ergodic rate under Rician fading is significantly higher than that under Rayleigh fading. It gradually decreases as the PA moves away from the favorable deployment region. More importantly, the analytical Rician results and the ideal LoS results almost completely overlap, which indicates that the contribution of the NLoS component to the achievable rate is negligible under a large Rician factor condition. The conventional practice of evaluating performance under ideal LoS conditions is justified. The similar phenomenon can be also observed in Fig.~\ref{fig: ER versus PA pos}(b), where PASS operates in the low-SNR regime but still with a large Rician factor. The Rayleigh ergodic rate in Fig.~\ref{fig: ER versus PA pos}(b) remains close to 0 for all PA positions, which indicates that the PASS performance becomes severely limited in the absence of a dominant LoS path in the low-SNR regime. In Fig.~\ref{fig: ER versus PA pos}(c), when the transmit power is low and the Rician factor is small, the ergodic rate gap between Rician and Rayleigh fading becomes smaller than that in Fig.~\ref{fig: ER versus PA pos}(b), since the strengthened NLoS scatterers make the Rayleigh and Rician channel gains less distinct. Meanwhile, the results under the ideal LoS condition falls noticeably below the Rician analytical and approximate results. This implies that the NLoS component can no longer be ignored, as it provides an additional power contribution to the ergodic rate.

\subsection{Comparison Between a ``Near'' PA and a ``Far'' PA}

PASS-enabled communication faces a fundamental tradeoff between spatial propagation loss and in-waveguide attenuation. As illustrated in Fig.~\ref{fig: nearPA_or_farPA scenario}(a), both a UE and an obstacle are located in the vicinity of the AP, and the UE is located closely behind the obstacle. Regarding the selection of PA position, on the one hand, a PA activated near the AP suffers only limited in-waveguide attenuation, but the corresponding link to the UE is restricted to NLoS propagation due to LoS blockage by the obstacle. On the other hand, a PA activated farther away along the waveguide can establish a LoS link by spatially bypassing the obstacle, but the in-waveguide attenuation increases. Such a tradeoff naturally raises the problem of which PA position can lead to superior performance.

To numerically demonstrate the impact of PA position on the performance, Fig.~\ref{fig: nearPA_or_farPA scenario}(a) illustrates a representative PASS scenario with physical LoS blockage, where the waveguide is partitioned into a LoS-feasible region and a blockage region. The obstacle's height is set to 3 m. The outage probability and ergodic rate versus PA position are illustrated in Fig.~\ref{fig: nearPA_or_farPA scenario}(b) and Fig.~\ref{fig: nearPA_or_farPA scenario}(c), respectively. Specifically, under the Rayleigh fading channel, a higher average NLoS power results in a greater ergodic rate and a lower outage probability. By contrast, under the Rician fading channel with a dominant LoS component, both the outage probability and the ergodic rate remain largely insensitive to the NLoS power and consistently outperform their Rayleigh counterparts. Therefore, several insights are highlighted: i) Maintaining LoS propagation supported by a far PA position is more effective than relying on NLoS propagation supported by a near PA position for guaranteeing both high rates and minimal outages, even if it results in a greater spatial propagation loss or in-waveguide attenuation; ii) A dominant LoS component renders the link robust to NLoS power variations; iii) When the LoS path is blocked, strong NLoS scattering can still support reliable communication.

\subsection{Optimal PA Position and Corresponding Performance}

\begin{figure}[t!]
  \centering
  \begin{minipage}{0.32\textwidth}
    \centering
    \includegraphics[width=0.9\textwidth]{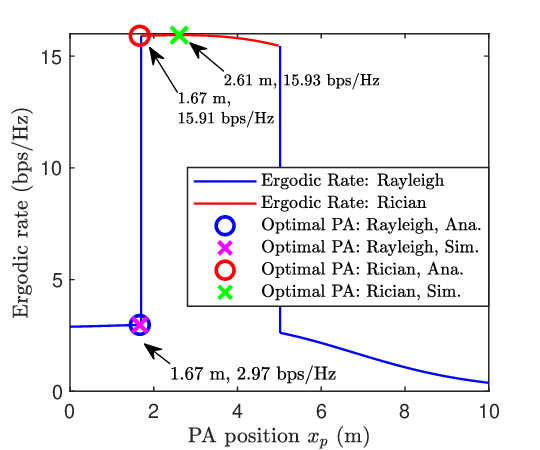}
    \centerline{\footnotesize(a)}
  \end{minipage}
  \begin{minipage}{0.48\textwidth}
    \centering
    \includegraphics[width=2.8 in]{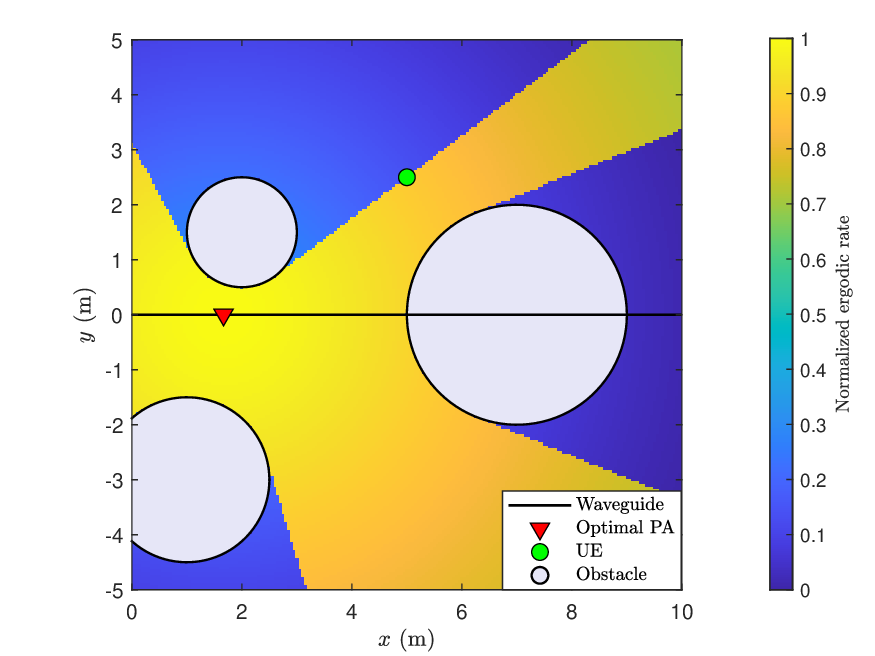}
    \centerline{\footnotesize(b)}
  \end{minipage}
  \caption{Performance characterization of a dispersed blockage scenario: (a) Ergodic rate versus PA position, where the analytical (Ana.) optimal PA positions are obtained from~\eqref{candidate optimal pos of PA Rayleigh} in the Rayleigh case and by solving~\eqref{problem of optimal PA position Rician} in the Rician case, while the simulations (Sim.) are obtained by maximizing the ergodic rate; (b) Normalized ergodic rate over the service area with the PA located at the analytical optimal position in the Rician case, where the blockage scenario is also shown.}
  \label{fig: Heatmap_Rician}
\end{figure}

Fig.~\ref{fig: Heatmap_Rician}(a) illustrates the ergodic rate of UE as a function of the PA's $x$-position along the waveguide in the scenario shown in Fig.~\ref{fig: Heatmap_Rician}(b). Primarily due to the position-dependent LoS blockage caused by several obstacles, a clear transition between the Rayleigh and Rician propagation regimes can be observed. The ergodic rate shows an abrupt increase at the first transition point (i.e., at the $x$-position of 1.67 m). Specifically, when the PA leaves the blockage region and enters the LoS-feasible region at around 1.67 m, the ergodic rate rises from 2.97 bps/Hz to about 15.9 bps/Hz, which reveals that a relatively small displacement of the PA along the waveguide may fundamentally alter the channel statistics in obstacle-rich environments.
The results of optimal PA position, as identified through analysis and simulation, further provide useful insights. In the Rayleigh case, the analytical and simulation results are almost identical, which validates the accuracy of the analytical characterization of~\eqref{problem of optimal PA position Rayleigh}. In the Rician case, the analytical and simulation-based optimal PA positions are different. Nevertheless, the ergodic rate achieved by these two PA positions are approximately identical. This is because the analytical solution in~\eqref{problem of optimal PA position Rician} is derived based on a maximum received SNR criterion, while the simulation-based optimum is obtained by directly maximizing the ergodic rate, which is also the original performance metric of interest. As stated by Jensen's inequality in~\eqref{ER Jensen to average SNR Rician}, these two criteria are not strictly equivalent. Although a mismatch in the resulting optimal PA positions between two criteria is expected in general, the proposed analytical criterion yields a solution with virtually no performance loss relative to the ergodic rate optimum while preserving tractability.

\begin{figure}[t!]
  \centering
  \begin{minipage}{0.32\textwidth}
    \centering
    \includegraphics[width=0.9\textwidth]{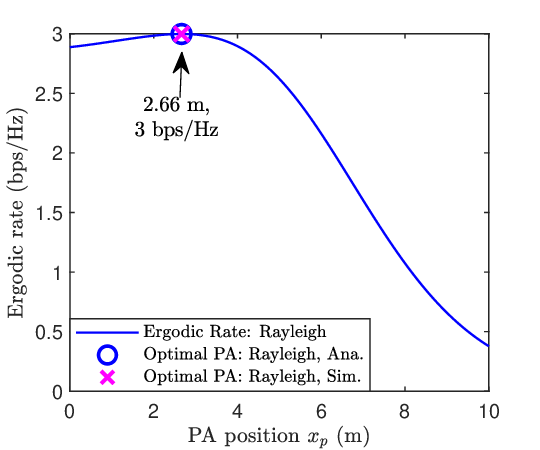}
    \centerline{\footnotesize(a)}
  \end{minipage}
  \begin{minipage}{0.48\textwidth}
    \centering
    \includegraphics[width=2.8 in]{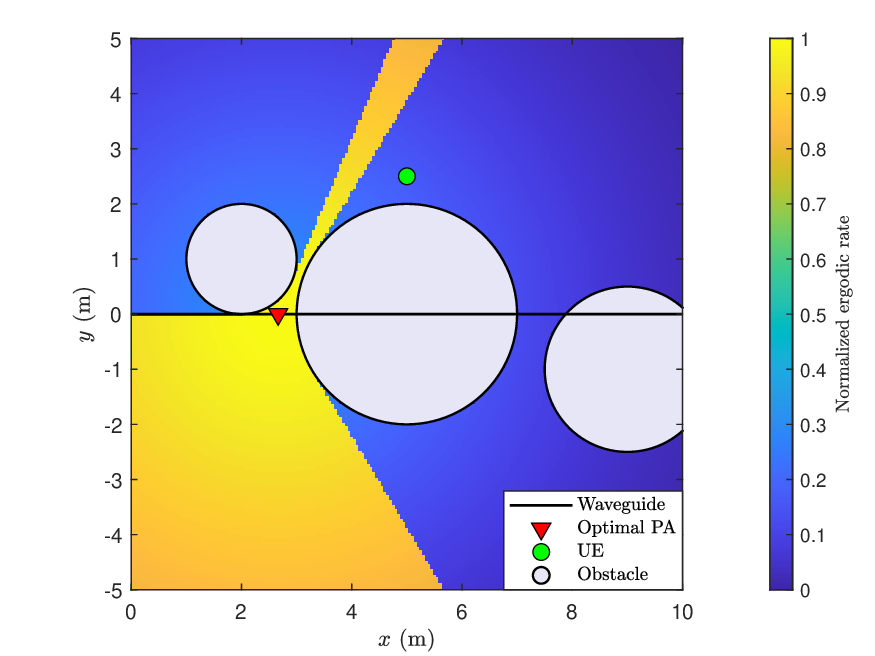}
    \centerline{\footnotesize(b)}
  \end{minipage}
  \caption{Performance characterization of a clustered blockage scenario: (a) Ergodic rate versus PA position, where the analytical (Ana.) optimal PA position is obtained by solving~\eqref{candidate optimal pos of PA Rayleigh} in the Rayleigh case, while the simulation (Sim.) is obtained by maximizing the ergodic rate; (b) Normalized ergodic rate over the service area with the PA located at the analytical optimal position in the Rayleigh case, where the blockage scenario is also shown.}
  \label{fig: Heatmap_Rayleigh}
\end{figure}

Fig.~\ref{fig: Heatmap_Rician}(b) illustrates the normalized ergodic rate distribution over the service area when the PA is fixed at the optimal analytical position corresponding to the Rician case. The high-rate region is not isotropically distributed around the PA. Instead, it is divided into ``sectors''-like areas blocked by obstacles, which indicates that the geometry and location of each obstacle directly determine the effective coverage pattern. Specifically, the high-rate regions correspond to the locations where favorable propagation conditions can be maintained relative, typically due to either a LoS path or a comparatively weak blockage effect. By contrast, the low-rate regions indicate severe degradation, which is primarily associated with LoS blockage and weak NLoS scattering. Taken together, Fig.~\ref{fig: Heatmap_Rician}(a) and Fig.~\ref{fig: Heatmap_Rician}(b) indicate that the blockage-aware ability is essential for realizing the full performance potential of PASS, especially in practical environments with irregular obstacles and intermittent LoS conditions.

Fig.~\ref{fig: Heatmap_Rayleigh}(a) illustrates the ergodic rate of UE as a function of the PA's $x$-position along the waveguide in the scenario shown in Fig.~\ref{fig: Heatmap_Rayleigh}(b), where obstacles completely block the LoS path for all feasible PA positions. As a result, the ergodic rate changes smoothly and remains at a much lower level. The maximum rate of approximately 3 bps/Hz is attained at around 2.66 m. It is observed that the analytical and simulation-based optimal PA positions for the Rayleigh case coincide exactly, which verifies the accuracy of \textbf{Proposition~\ref{proposition: candidate optimal pos of PA Rayleigh}}.

Fig.~\ref{fig: Heatmap_Rayleigh}(b) illustrates the normalized ergodic rate distribution over the service area when the PA is fixed at the optimal analytical position corresponding to the Rayleigh case. Compared to the scenario in Fig.~\ref{fig: Heatmap_Rician}(b), the high-rate region in Fig.~\ref{fig: Heatmap_Rayleigh}(b) is substantially compressed, while the low-rate regions around the UE become more dominant. It is worth noting that the high-rate region is not simply interpreted as a LoS-supported coverage zone, but as the least unfavorable region within a densely obstructed environment. From the perspective of system design, the deployment strategy for waveguides, including their locations and quantity, significantly impacts the performance of PASS, particularly in complex environments. If the deployment strategy for waveguides does not align with the geometry of service environment, it is hard to fully leverage the performance gains achieved through the flexible activation of PAs.

\section{Conclusion}
In this paper, the performance of PASS was investigated by jointly considering LoS blockage and NLoS propagation. A geometry-aware blockage model was adopted to link the PASS channel model, where the blockage region on the waveguide was defined. The channel models were developed by jointly accounting for spatial propagation loss and in-waveguide attenuation, where the latter consisted of both LoS and NLoS components. A single-PA single-UE scenario was studied under both Rayleigh and Rician fading channels, where the analytical expressions for the outage probability and ergodic rate were derived. The deployment criteria for optimal PA were provided and discussed.

Several directions could be pursued in future work. For analytical tractability, it was assumed that the geometries and locations of obstacles were deterministic, so that LoS blockage could be characterized in a deterministic manner. In practical environments, however, obstacles might be randomly distributed in terms of shape and position, and they might even be dynamic. Therefore, more general blockage models could be incorporated. Furthermore, the analysis could be extended from single-PA and single-UE scenarios to multi-PA and multi-user scenarios, where multiple access schemes and inter-user interference management could be jointly considered. Finally, multi-waveguide deployments could be conducted, including resource allocation and beamforming.

\section*{Appendix~A \\ Proof of Corollary~\ref{coro: ER Rician approximate}}\label{Appendix:As}
\renewcommand{\theequation}{A.\arabic{equation}}
\setcounter{equation}{0}
The random variable $\left| {h_L} + {h_N} \right|^2$ can be expressed as follows:
\begin{equation}\label{channel gain 1}
\begin{split}
Z & \triangleq {\left| h_L + h_N \right|^2} \\
     & = {\left| h_L \right|^2} + 2\operatorname{Re} \left\{ {h_L^ * {h_N}} \right\} + {\left| h_N \right|^2},
\end{split}  
\end{equation}
where $\left(  \cdot  \right)^ * $ and $\operatorname{Re} \left\{ \cdot \right\}$ denote the conjugate and the real part of a complex number, respectively. Note that $2\operatorname{Re} \left\{ h_L^ * {h_N} \right\}$ follows a complex Gaussian distribution, i.e.,
\begin{equation}\label{channel gain 2}
2\operatorname{Re} \left\{ h_L^ * {h_N} \right\} \sim \mathcal{CN}\left( {0,2{\left| h_L \right|}^2 \sigma _N^2} \right).
\end{equation}

Thus, the expectation of $Z$ is given by
\begin{equation}\label{expectation of X}
\begin{split}
\mathbb{E}\left[ Z \right] & = {\left| h_L \right|^2} + 2\operatorname{Re} \left\{ {h_L^ * \mathbb{E}\left[ {{h_N}} \right]} \right\} + \mathbb{E}\left[ {\left| {{h_N}} \right|}^2 \right] \\
     & = \frac{\eta }{d^2} + \frac{\mu _N^2}{d^{\alpha _N}},
\end{split}
\end{equation}
and the variance of $X$ is given by
\begin{equation}\label{variance of X}
\begin{split}
\operatorname{Var}\left[ Z \right] & = 2{\left| h_L \right|^2}\sigma _N^2 + \sigma _N^4 \\
     & = \frac{2\eta \mu _N^2}{d^{{\alpha _N} + 2}} + \frac{\mu _N^4}{d^{2{\alpha _N}}},
\end{split}
\end{equation}
where $\operatorname{Var}\left[ \cdot \right]$ denotes the variance.

We set ${f_1}\left( x \right) = \log _2 \left( {1 + \rho x} \right)$. By expanding ${f_1}\left( x \right)$ to a Taylor series at $x = \mathbb{E}\left[ Z \right]$~\cite{Table_of_integrals}, the expectation $\mathbb{E}\left[ f_1\left( Z \right) \right]$ can be expressed as follows:
\begin{equation}\label{taylor expansion of ergodic rate L 1}
\mathbb{E}\left[ {f_1\left( Z \right)} \right] \approx {f_1}\left( {\mathbb{E}\left[ Z \right]} \right) + \frac{{f_1^{\prime \prime}\left( {\mathbb{E}\left[ Z \right]} \right)}}{2}\operatorname{Var} \left[ Z \right],
\end{equation}
where the second derivation of ${f_1}\left( x \right)$ is given by
\begin{equation}\label{f 1 prime prime}
f_1^{\prime \prime} \left( x \right) =  - \frac{\rho ^2}{{{\left( {1 + \rho x} \right)}^2}\ln 2}.
\end{equation}

According to~\eqref{taylor expansion of ergodic rate L 1}, the ergodic rate in~\eqref{ER Rician 1} can be approximated as follows:
\begin{equation}\label{taylor expansion of ergodic rate L 2}
R_L \approx {\log _2}\left( {1 + \rho \mathbb{E}\left[ Z \right]} \right) - \frac{{\rho ^2}}{2{{\left( {1 + \rho \mathbb{E}\left[ Z \right]} \right)}^2}\ln 2}\operatorname{Var} \left[ Z \right].
\end{equation}

Substituting~\eqref{expectation of X} and~\eqref{variance of X} into~\eqref{taylor expansion of ergodic rate L 2} yields~\eqref{ER Rician approximate}. Hence, the proof is complete.

\section*{Appendix~B \\ Proof of Proposition~\ref{proposition: Rician Rate greater than LoS Rate}}\label{Appendix:Bs}
\renewcommand{\theequation}{B.\arabic{equation}}
\setcounter{equation}{0}
We set $\Phi \left( z \right) = {\log _2}\left( {1 + \rho {{\left| z \right|}^2}} \right)$, where $z \in \mathbb{C}$. The ergodic rate in~\eqref{ER Rician 1} under a Rician fading channel can be expressed as follows:
\begin{equation}\label{Rate Rician integrated}
{R_L} = \mathbb{E}\left[ {\Phi \left( h_L + h_N \right)} \right],
\end{equation}
and the achievable rate under LoS propagation can be expressed as follows:
\begin{equation}\label{Rate pure LoS}
{\hat R}_L = {\log _2}\left( {1 + \rho {\left| h_L \right|}^2} \right) = \Phi \left( h_L \right).
\end{equation}

The Laplacian of $\Phi \left( z \right)$ is given by
\begin{equation}\label{Laplacian of subharmonic function}
{\nabla ^2}\Phi \left( z \right) = \frac{4\rho }{{{\left( {1 + \rho {{\left| z \right|}^2}} \right)}^2}\ln 2},
\end{equation}
where $\nabla ^2$ denotes the Laplace operator. Since $\rho > 0$, ${\nabla ^2}\Phi \left( z \right) > 0$, then $\Phi \left( z \right)$ is a strictly subharmonic function~\cite{Ransford_1995}.

Let $h_N = \Gamma {e^{j\theta }}$, where $\Gamma  = \left| h_N \right|$ and $\theta$ is the argument of $h_N$.
Since $h_N$ in~\eqref{NLoS component Gaussian} is a circularly symmetric complex Gaussian random variable, $\theta$ is uniformly distributed over $\left[0, 2\pi \right)$, and it is independent of $\Gamma$. For any fixed $\Gamma_0 > 0$, the following conditional expectation holds true:
\begin{equation}\label{conditional expectation of Rician channel gain}
\mathbb{E}\left[ \Phi \left( h_L + h_N \right)\,\middle|\,\Gamma  = \Gamma _0 \right] = \frac{1}{2\pi }\int_0^{2\pi } {\Phi \left( h_L + {\Gamma _0}{e^{j\theta }} \right)d\theta }.
\end{equation}

Since $\Phi \left( z \right)$ is strictly subharmonic, the mean inequality for subharmonic functions yields~\cite{Ransford_1995}
\begin{equation}\label{mean inequality for subharmonic functions 1}
\frac{1}{2\pi }\int_0^{2\pi } {\Phi \left( {h_L + {\Gamma _0}{e^{j\theta }}} \right)d\theta }  > \Phi \left( h_L \right).
\end{equation}

Thus, for all $\Gamma_0 > 0$, the following inequality holds true:
\begin{equation}\label{mean inequality for subharmonic functions 2}
\mathbb{E}\left[ \Phi \left( h_L + h_N \right)\,\middle|\,\Gamma  = \Gamma _0 \right] > \Phi \left( h_L \right),
\end{equation}
which further implies
\begin{equation}\label{expectation of conditional expectation}
\mathbb{E}\left[ \Phi \left( h_L + h_N \right) \right] = \mathbb{E}\left[ \mathbb{E}\left[ \Phi \left( {h_L} + {h_N} \right)\,\middle|\,\Gamma \right] \right] > \Phi \left( h_L \right).
\end{equation}

Hence, $R_L > \hat{R}_L$ holds true. The proof is complete.

\section*{Appendix~C \\ Proof of Lemma~\ref{lemma: the intersections pf blockage region}}\label{Appendix:Cs}
\renewcommand{\theequation}{C.\arabic{equation}}
\setcounter{equation}{0}
To determine the endpoints of the blockage region on the waveguide, it is necessary to determine two tangent lines to the cross-section of each obstacle, both of which pass through the UE's position. The tangent equation of the obstacle that passes through the UE's position is defined as follows:
\begin{equation}\label{tangent def}
y - {y_u} = m\left( {x - {x_u}} \right),
\end{equation}
where $m$ denotes the slope.

On the $x$-$y$ plane, the distance from the center of the $n$-th obstacle to the tangent line is given by
\begin{equation}\label{the distance from obs center to tangents}
\frac{\left| {m\left( {{x_n} - {x_u}} \right) - \left( {{y_n} - {y_u}} \right)} \right|}{\sqrt {{m^2} + 1} } = r_n.
\end{equation}

The slope of the tangent line can be obtained by treating it as an independent variable in~\eqref{the distance from obs center to tangents}. Thus,~\eqref{the distance from obs center to tangents} can be written as a quadratic equation as follows:
\begin{equation}\label{the equation of tangents' slope}
\left( {{\left( {\Delta {x_n}} \right)}^2} - r_n^2 \right){m^2} - 2\Delta {x_n}\Delta {y_n}m + \left( \Delta {y_n} \right)^2 - r_n^2 = 0,
\end{equation}
where $\Delta {x_n} = {x_n} - {x_u}$ and $\Delta {y_n} = {y_n} - {y_u}$.

In general, since the position of the UE is outside the obstacle, i.e., ${\left( {\Delta {x_n}} \right)}^2 + {\left( {\Delta {y_n}} \right)}^2 > r_n^2$, the discriminant of~\eqref{the equation of tangents' slope} satisfies $4r_n^2\left( {\left( \Delta {x_n} \right)}^2 + {\left( \Delta {y_n} \right)}^2 - r_n^2 \right) > 0$. Therefore, if $\Delta {x_n} \ne r_n$, two real roots of slopes can be obtained by solving~\eqref{the equation of tangents' slope}, which are expressed as
\begin{equation}\label{tangents' slope general 1}
m_1 = \frac{{\Delta {x_n}\Delta {y_n} + {r_n}\sqrt {{{\left( {\Delta {x_n}} \right)}^2} + {{\left( {\Delta {y_n}} \right)}^2} - r_n^2} }}{{{\left( {\Delta {x_n}} \right)}^2} - r_n^2},
\end{equation}
and
\begin{equation}\label{tangents' slope general 2}
m_2 = \frac{{\Delta {x_n}\Delta {y_n} - {r_n}\sqrt {{{\left( {\Delta {x_n}} \right)}^2} + {{\left( {\Delta {y_n}} \right)}^2} - r_n^2} }}{{{\left( {\Delta {x_n}} \right)}^2} - r_n^2},
\end{equation}
respectively. If $\Delta {x_n} = {r_n}$, then there is a special case in which only one slope exists, because one of the tangent lines is perpendicular to the $x$-axis. Thus,~\eqref{the equation of tangents' slope} is simplified as $ - 2 r_n \Delta {y_n}m + {\left( {\Delta {y_n}} \right)^2} - r_n^2 = 0$, where the slope is expressed as follows:
\begin{equation}\label{tangents' slope special}
m_3 = \frac{{{\left( {\Delta {y_n}} \right)}^2} - r_n^2}{2 r_n \Delta {y_n}}.
\end{equation}

The intersections of the two tangent lines with the waveguide line are obtained by substituting $y=0$ into~\eqref{tangent def}, which are expressed as
\begin{equation}\label{initial intersection 1}
x_n^{b - }  =
\begin{cases}
\min \left\{ {x_u} , {x_u} - \frac{y_u}{m_3} \right\}, & \mbox{if } \Delta {x_n} = {r_n}, \\
\min \left\{ {x_u} - \frac{y_u}{m_1}, {x_u} - \frac{y_u}{m_2} \right\}, & \mbox{otherwise},
\end{cases}
\end{equation}
and
\begin{equation}\label{initial intersection 2}
x_n^{b + }  =
\begin{cases}
\max \left\{ {x_u} , {x_u} - \frac{y_u}{m_3} \right\}, & \mbox{if } \Delta {x_n} = {r_n}, \\
\max \left\{ {x_u} - \frac{y_u}{m_1}, {x_u} - \frac{y_u}{m_2} \right\}, & \mbox{otherwise},
\end{cases}
\end{equation}
respectively.

Recall that the PA moves within the finite domain of the waveguide, i.e., $x_p \in \left[ 0, D_L \right]$, the two endpoints of the blockage region are respectively expressed as
\begin{equation}\label{endpoints of blockage region 1}
x_n^ -  = \max \left\{ 0,x_n^{b - } \right\},
\end{equation}
and
\begin{equation}\label{endpoints of blockage region 2}
x_n^ +  = \min \left\{ {D_L},x_n^{b + } \right\},
\end{equation}
respectively. Hence, the proof is complete.

\section*{Appendix~D \\ Proof of Proposition~\ref{proposition: candidate optimal pos of PA Rayleigh}}\label{Appendix:Ds}
\renewcommand{\theequation}{D.\arabic{equation}}
\setcounter{equation}{0}
Regarding the problem in~\eqref{problem of optimal PA position Rayleigh}, a function is defined as follows:
\begin{equation}\label{f_2(x)}
{f_2}\left( x_p \right) = {\left( {{{\left( {x_u - x_p} \right)}^2} + d_0^2} \right)^{\frac{\alpha _N}{2}}}{e^{2{\alpha _g} x_p}},
\end{equation}
which denotes the product of spatial propagation loss and in-waveguide attenuation. Thus, maximizing the total channel gain is equivalent to minimizing $f_2 \left( x_p \right)$, and the local minimum point of $f_2 \left( x_p \right)$ is a candidate of optimal PA position.

Since $\alpha _N$, $\alpha _g$, and $d_0 > 0$, the logarithm of ${f_2}\left( x_p \right)$ can be taken as follows:
\begin{equation}\label{f_3(x)}
{f_3}\left( x_p \right) = 2{\alpha _g} x_p + \frac{\alpha _N}{2}\ln \left( {{{\left( {x_u - x_p} \right)}^2} + d_0^2} \right).
\end{equation}

The first-order derivative of $f_3 \left( x_p \right)$ is given by
\begin{equation}\label{f_3_prime(x)}
{f_3^{\prime}}\left( x_p \right) = 2{\alpha _g} - \frac{{{\alpha _N}\left( {x_u - x_p} \right)}}{{{{\left( {x_u - x_p} \right)}^2} + d_0^2}}.
\end{equation}

To find the stationary points at which $f_3 \left( x_p \right)$ attains its minimum value, it is essential to identify the real roots of its first derivative when equal to zero, i.e., solve the equation as follows:
\begin{equation}\label{f_3_prime(x) = 0}
2{\alpha _g}{\left( {x_p - x_u} \right)^2} + {\alpha _N}\left( {x_p - x_u} \right) + 2{\alpha _g}d_0^2 = 0.
\end{equation}

For the quadratic equation ${f_3^{\prime}}\left( x_p \right) = 0$, its discriminant is $\alpha _N^2 - 4\alpha _g^2d_0^2$. Note that if ${d_0} > \tfrac{1}{4{\alpha _g \alpha _N}}$, there is no real root to the equation ${f_3^{\prime}}\left( x_p \right) = 0$. If ${d_0} = \tfrac{1}{4{\alpha _g \alpha _N}}$, there are two identical real root to the equation ${f_3^{\prime}}\left( x_p \right) = 0$. Thus, if ${d_0} \geqslant \tfrac{1}{4{\alpha _g \alpha _N}}$, ${f_3^{\prime}}\left( x_p \right) \geqslant 0$ holds true, which means that ${f_3}\left( x_p \right)$ is monotonically increasing. Conversely, if ${d_0} < \tfrac{1}{4{\alpha _g \alpha _N}}$, there are two distinct real roots to the equation ${f_3^{\prime}}\left( x_p \right) = 0$, which are given by
\begin{align}
\label{f_3_prime(x) root 1}
x_{p1} & = {x_u} - \frac{{\alpha _N} + \sqrt {\alpha _N^2 - 16\alpha _g^2d_0^2} }{4{\alpha _g}}, \\
\label{f_3_prime(x) root 2}
x_{p2} & = {x_u} - \frac{{\alpha _N} - \sqrt {\alpha _N^2 - 16\alpha _g^2d_0^2} }{4{\alpha _g}}.
\end{align}

Since $x_{p1} < x_{p2}$ holds, ${f_2}$, $x_{p1}$ and $x_{p2}$ are the local maximum and minimum point of ${f_3}\left( x_p \right)$, respectively. They are also the local maximum and minimum points of ${f_2}\left( x_p \right)$, respectively.

In addition, the optimal PA position is also determined based on whether $x_{p1}$ and $x_{p2}$ are within the interval of the blockage region. Recall that the blockage region is defined as $\left[ x_n^-, x_n^+ \right]$, there are three possible cases of the optimal PA position: 1) If $x_n^- \leqslant x_{p2} \leqslant x_n^+$, which means that the minimum point of ${f_2}\left( x_p \right)$ is within the blockage region, the optimal PA position is $x_{p2}$; 2) If $x_n^+ < x_{p2}$, the optimal PA position takes the minimum point between ${f_2}\left( x_n^- \right)$ and ${f_2}\left( x_n^+ \right)$; 3) If $x_n^- > x_{p2}$ or $x_n^+ < x_{p1}$, the optimal PA position is $x_n^-$. Hence, all possible candidate optimal PA positions are summarized in~\eqref{candidate optimal pos of PA Rayleigh}, and the proof is complete.

\bibliographystyle{IEEEtran}
\bibliography{IEEEabrv,Blockage_LoS_NLoS_PASS}

\end{document}